# Size-Independent Non-Equilibrium Fluctuations

Maria K. Koleva
Institute of Catalysis, Bulgarian Academy of Science
1113 Sofia, Bulgaria
e-mail: mkoleva@bas.bg

## Abstract

A local quantum phenomenon that gives rise to generic for all surface reactions macroscopic fluctuations is studied. The issue is viewed with respect to the necessary conditions for a long-term stable evolution of any natural and artificial system. It is shown that global coupling of the local fluctuations is necessary for providing a long-term stability of the system. A successful coupling mechanism is achieved on the grounds of new assumptions about the Hamiltonian response to certain perturbations. The coupling mechanism acts towards a global synchronisation, i.e. to a coherent response of the excited species to any further perturbation. It is proven that the synchronisation is a scale-free process that has universal properties, *e.g.* it is insensitive to the chemical identity of the reacting species and to the particularities of the surface reaction. Its hallmark is that the global adsorption rate exhibits permanent temporal variations whose amplitude is independent of the system size.

The presence of these fluctuations fundamentally changes the temporal behavior of the system, namely it becomes pulse-like both on the quantum and the macro-level. The pulse-like behavior gives rise to a persistent continuous band at the quantum spectra whose major properties are: (I) it does not correspond to any real radiation; (ii) its presence is insensitive to the particularities of the system and the incident radiation; (iii) its shape and the infrared edge are typical for the $1/f^{\alpha}$-type noise. These properties give rise to its name: alias $1/f^{\alpha}$-type noise.

## 1. Introduction

Recently, a general condition that relates the long-term stability of any natural and artificial system and exerting fluctuations has been put forth [1-3]. It asserts that a system stays stable if and only if the amount and the rate of exchange of energy and/or matter currently involved in any transition does not exceed the thresholds of stability of the system. Thus, both locally and globally the fluctuations should permanently be sustained bounded.

However, when only short range interactions are involved in the elementary processes, the local dynamics in the extended systems is non-correlated both in space and time. This immediately renders developing of the local defects such as strain, overheating, sintering etc. At the next instant the spatial configuration changes and the local defects move. Due time course their interaction produces local reconstruction, creates mechanical, thermal and/or other defects. Eventually the process yields the system breakdown.

So, apparently the long-term stability calls for persistent coupling of the local fluctuations so that to sustain their amplitude permanently bounded both locally and globally.

A wide spectrum of works aims to explore an effect of correlations of fluctuations in extended systems as an interplay among noise correlations, non-linearity and spatial coupling. Yet, the stochastic variables and noise sources in all the developed so far approaches are modelled by the use of a Wiener process whose increments are independent and unbounded.

Thus, though a cooperation of the fluctuations is available, the sequence of spatio-temporal configurations through which it arrives to the global coupling varies in uncontrolled way that is irreconcilable with the idea of the boundedness.

It is to be expected that the boundedness imposes certain general constraints on the structure of the state space. In turn, that gives rise to the question whether the state space is connected by a Markovian chain that appears to be an invariant measure. If so, we can apply the developed so far approaches such as Fokker-Planck equation, master equation etc. to our problem. Otherwise, we should look for new approach(es) and to anticipate new properties.

One immediate outcome of the relation boundedness-stability is that each point in the state space selects a set of admissible transitions such that they do not violate the stability of the system. To compare, the overlook of that relation admits transitions between *any* two states. Furthermore, the boundedness renders the state space of a stable system always confined in a finite volume. These properties ensure the enumeration of the state space so that all the accessible states to a given one are the nearest neighbors.

Now, let us suppose that the transition from the state $j$ to the neighbor state $i$ depends only on whether the transition to $j$ has happened. However, the transition to $j$ depends on whether it comes from its nearest neighbourhood $k$. Hence, the transition to $i$ is set on the chain of the previous transitions $...l...kj$. So, is our process non-Markovian though the Chapman-Kolmogorov relation holds?! On the one hand, it seems Markovian because the transition from $j$ to $i$ depends only on $j$. On the other hand, it is non-Markovian, because any admissible transition depends on the succession of the previous ones. (Examples of non-Markovian chains that fulfil Chapman-Kolmogorov relation are presented in [4]. The question now becomes whether the "non-Markovian" succession covers the whole story of the process. The clue is in the Lindeberg theorem [4] that states: any bounded irregular sequence (BIS) has finite mean and finite variance. Then, our trajectory being confined in a finite volume is a BIS. So, it "loses" its memory any time it crosses the mean. Though any "walk" that starts at the mean comes back to it in a finite time, the size of the "walks" varies up to the size of the state space. It means that a non-Markovian "walk" can comprise most of the states.

As a result, we should look for a new approach to establish the properties of the state space. Indeed, it has been found out [3] that the coarse-grained structure of a state space subject to the constraints imposed by the boundedness exhibits strong chaotic properties insensitive to the particularities of the system.

Yet, when the evolution of an extended many-body system is developed under local rules, the lack of correlation among the local spatio-temporal fluctuations renders a self-sustained boundedness rather occasional and specific to a system and/or to certain states. This gives rise to the question about a mechanism that give rise to boundedness at each point of the state space.

The task of the present paper is to propose a general mechanism for coupling spatio-temporal fluctuations at surface reactions that proceeds at infinite interfaces gas/solid. It is proven that the mechanism is insensitive to the particularities of the reaction mechanism, interface properties etc. As a result of the coupling, at any value of the control parameters the system exerts macroscopic fluctuations whose amplitude is bounded. The important property of this coupling mechanism is its persistence, i.e. for each state of the system it determines a non-zero set of states accessible from that one. In turn, this provides the "continuity" of the state space trajectories.

A prominent hallmark of the entire concept is that the fluctuations amplitude is insensitive to the system size. This in sharp contrast to the strong size-dependence of the macroscopic fluctuations resulted by the developed so far approaches.

The major dilemma is whether the coupling mechanism drives the local fluctuations "dying down" or there exists a coupling mechanism so that the local fluctuations become coherent? The first alternative renders the macroscopic fluctuations available only if long-range interactions are involved or at phase transitions. The second alternative requires a mechanism of coupling local fluctuations available at each value of the control parameters, reaction mechanism, system size and properties etc.

The first step is to present a new type of local fluctuations generic for all surface reactions. The task is to clarify that their properties make the first alternative non realistic. They has been introduced recently [1-2] and named diffusion-induced noise. Their significance is set on their ubiquity and insensitivity to the particularities of the surface reaction.

The driving mechanism of the diffusion-induced noise is presented for the adsorption, since it is a step prerequisite of any surface reaction at any control parameter choice. The mechanism is based on the interplay between: (I) the lack of correlation between moments and points of the gas phase species trapping at the interface; (ii) the generic property of any adlayer that no more than one species can be adsorbed at a single active site. That interplay causes fundamental changes of the properties of the overall probability for adsorption (correspondingly the adsorption rate). Given is a species trapped in a vacant site. Its further relaxation to the ground state can be interrupted by an adspecies that arrives at the same site and most probably occupies it. Thus the adspecies violates the further trapped species relaxation at that site, since no more than one species can be adsorbed at a single site. The trapped species can complete the adsorption if and only if after migration it finds another vacant site. The impact of the adspecies intervention to the trapped species probability for adsorption is twofold: first, it cannot be considered as a perturbation, since it changes the adsorption potential qualitatively, namely from attractive it becomes repulsive. That is why, that type of interaction has been called diffusion-induced non-perturbative interaction. Second, the lack of coherence between the trapping moment and the moment of adspecies arrival makes the probability for adsorption multi-valued function: each selection corresponds to a certain level of relaxation at which a diffusion-induced non-perturbative interaction happens. Therefore, the adspecies mobility brings about fundamental duality of the probability for adsorption (and of the adsorption rate correspondingly): though each selection can be computed by an appropriate quantum-mechanical approach, the establishing of a given selection is a stochastic process since it is a random choice of a single selection among all available. Since the diffusion-induced non-perturbative interactions are local events, the non-correlated mobility of the adspecies produces a lack of correlation between the established selections at any distance and at any instant. As a result, the produced adlayer would be always spatially non-homogeneous even in the academic case of identical adsorption and mobility properties of all types of adspecies. Outlining, the non-correlated diffusion-induced non-perturbative interactions always make the adsorption rates that come from different adsorption sites non-identical that immediately produces spatial non-homogeneity. Furthermore, the induced non-homogeneity would be permanently sustained by the lack of coherence between the trapping moments and the adspecies mobility. In turn, the adlayer configuration would vary in uncontrolled way so that in a short time it would cause either the reaction termination or the system breakdown. Thus, a stable long-term evolution is available if and only if there is a mechanism that suppresses the induced non-homogeneity.

It is obvious that the presence of the diffusion-induced noise is insensitive to the details of the adsorption Hamiltonian, to the chemical identity of the adspecies and the particularities of the interface. Further, it exists at each value of the control parameters. So, the diffusion-induced noise persists at each point of the state space and is generic for all surface reactions.

The supposition about a mechanism of fluctuations "dying down" is non realistic because there is permanent producing of local fluctuations - the gas phase bombardment of the surface is enduring and insensitive to what happens there. Thus, the only way out is to look for a mechanism that removes the induced non-homogeneity through a feedback that couples the local fluctuations.

A successful coupling mechanism needs a feedback that acts toward evening of the initially non-identical adsorption rates through making the coupled species "response" to further perturbation(s) coherent. It should be grounded on a strong coupling adlayer-interface, namely: the energy of colliding species dissipates to local cooperative excitations of the interface. In turn, the impact of these local modes on the colliding species is supposed large enough to induce a new transition that dissipates through the excitation of another local cooperative modes and so on. The feedback ceases its action whenever the colliding species response becomes coherent.

However, the used so far weak coupling approach renders any feedback set on the interaction adspecies-interface local and non-correlated both in space and time. It provides only local evening of the current states of the colliding species. Suppose that the collision energy dissipates through the excitement of local cooperative modes. The weak-coupling approach considers any change of the adsorption Hamiltonian due to these excitations a perturbation. Consequently, the latter cannot produce a large enough change of the species state to induce a new transition. As a result, the interaction colliding species-interface stops. Since only short range interactions are considered, a new transition happens only when other species collide brought together by their mobility.

To make the required feedback available we introduce two new major assumptions. The first one is that any adsorption Hamiltonian separates into a "rigid" and a "flexible" part under the disturbances induced by the non-correlated mobility of the adspecies. The "rigid" part is specific to the system and comprises the low-lying levels where the adspecies impact can be treated as perturbation. The "flexible" part corresponds to that highly excited states where the impact of the mobile adspecies is as strong as potentials that create the Hamiltonian. So, the "flexibility" of the Hamiltonian implies that it always changes under the non-correlated movements. The sensitivity to the corresponding environments renders the "flexible" parts of the Hamiltonians that comes from different adsorption sites non-identical.

The separation of the adsorption Hamiltonians into a "rigid" and "flexible" part is another manifestation of our general concept about the boundedness of energy and/or matter involved in any process. In this particular case, it implies that the energy and matter involved in the creation of any Hamiltonian is finite. Correspondingly, it can be destroyed by finite amount of energy.

The next major assumption is about the general properties of the induced transitions. It is supposed that they are non-radiative and dissipate always through excitation of appropriate local low-frequency cooperative modes, e.g. acoustic phonons. The new point is that the local low-frequency excitations participate to the "flexible" Hamiltonian the same way as the mobile adspecies. In turn this constitutes a non-perturbative feedback, namely: the mobile environment of an excited species induces a non-radiative transition that dissipate through excitation of local acoustic phonon(s) (or other appropriate excitation(s)). The latter causes an immediate non-perturbative change of the "flexible" part of the Hamiltonian that in turn induces a new transition. The feedback acts toward evening of the initially non-identical adsorption rates via "synchronisation" of the initially non-identical "flexible" Hamiltonians so that they "respond" coherently to further perturbation(s). After the synchronisation is completed, the further relaxation proceeds through the "rigid" part of the Hamiltonian.

In the next section it is proven that though the "flexible" parts of the Hamiltonians are not identical, they have chaotic properties. Following the Bohigas et al conjecture [5-6], their spectra share the same distribution of the nearest level spacing, namely the Wigner one. It provides the insensitivity of the feedback to the chemical identity of the reacting species since the Wigner distribution involves a single parameter to characterise the chaotic states and the transitions between them.

Next, it is shown that the collisions between the species in chaotic states, i.e. the weakest perturbations that drive the feedback, dissipate through the excitation of acoustic phonons. The particular property of the feedback based on dissipation through acoustic phonons is that the coupling area (the wave-length) enlarges with the decrease of the energy of a transition that drives the feedback (frequency). In turn, this provides the globalisation of the synchronisation through the enlargement of the coupling area.

The universality of the feedback chaotic states $\Leftrightarrow$ acoustic phonons is set on the insensitivity of the chaotic spectrum to the chemical identity of the reacting species (a chaotic spectrum is characterised by a single parameter) and the insensitivity of the acoustic phonons to the details of the adspecies configuration and the surface itself (its dispersion relation involves a single parameter - sound velocity).

Further, it is proven that the global adsorption rate does not depend on the details of the spatio-temporal configuration of the "chaotic" species and the adspecies configuration. To the most surprise, it is always identical to the individual adsorption rate that comes from certain local configuration. It is verified that the feedback always selects that individual adsorption rate which initially is in the most favorable local configuration: such that the difference in the state of that species and its immediate neighbors is the smallest. It is proven as well that the synchronisation is a scale-free process that does not blur the individual properties of the "most favourable" adsorption rate.

The most pronounced differences among the individual adsorption rates come from the undergoing of a diffusion-induced non-perturbative interaction. Thus, the particularity of the "most favourable" adsorption rate is determined by the established selection. Furthermore, the lack of correlation between the trapping moments and the adspecies mobility makes all the selections equiprobable. Therefore, at next instant the "most favorable" adsorption rate involves another selection of a diffusion-induced non-perturbative interaction. Consequently, the random choice of a single selection among all available brings about irregular variations of the global adsorption rate in the course of time.

The third section elucidates the insensitivity to the system size of the macroscopic fluctuations created by the global adsorption rate temporal variations. The current macroscopic fluctuation size is the product of the current global adsorption rate and the surface area. The size of any adsorption rate is determined by the intensity of the diffusion-induced non-perturbative interactions that is proportional to the current coverage of adspecies. Thus, the current macroscopic fluctuation size is proportional to the current adspecies number. It should be stressed that the amount of species that the surface can adsorb is limited not by its size but by its saturation threshold. It, however, is set on the thresholds of the stability of the system that fix the limits of the range of energy and/or matter fluctuations so that the system stays permanently stable. This immediately renders the size-independence of the fluctuation amplitude. The boundedness of the considered fluctuations is ensured: locally by the property that only one species can be adsorbed at a single site, globally by the presence of saturation threshold.

The hallmarks of the variations of the adsorption rate are their obvious boundedness, persistence at any point of the state space and the duality determinism-stochastisity. The latter implies that though each selection can be computed by an appropriated quantum approach, the establishing of a single selection is a stochastic process since it is random choice of a single selection among all available. However, the most prominent property of these variations is that they come from a coupling process. Thus, the current variation of the adsorption rate is the same for each of the adspecies.

As it was mentioned already we have introduced the idea that stable evolution of any natural system is sustained if and only if the energy and matter fluctuations are bounded so that the system permanently stays within the thresholds of stability [1-3]. So, the hallmark of any stable evolution is the boundedness of a time series that depicts its temporal behaviour. It has been established [2] that an irregular bounded sequence (BIS) has certain common properties called $1/f^\alpha$-type noise. Hence, the considered in the present paper inevitable macroscopic fluctuations should be regarded as the origin of the $1/f^\alpha$-type noise behavior at the surface reactions. Indeed, the $1/f^\alpha$-type noise behavior has been established at the catalytic oxidation of $HCOOH$ over supported $Pd$ catalyst [7].

It should be stressed that the idea that self-organization of the local fluctuations brings about $1/f$ behavior is not new. It has been introduced by Bak, Tang and Wiesenfeld [8] who suggested that natural systems might organize themselves to a critical state through their local dynamics. They named this kind of organization to criticality "self-organized criticality" (SOC). This behavior is usually characterized by a power law distribution function of the activity (avalanches) in the critical dynamical systems and by existence of scale invariance of the distribution function. Bak, Tang and Wiesenfeld introduced SOC in order to explain the $1/f$ tails in the power spectra of the extended many-body systems. Later it was proven [9-10] that the model is not coherent with $1/f$ behavior but with the $1/f^2$ one. Despite its elegance SOC cannot display a universal power law exponent. Another important disadvantages of the model are: the strong dependence of the power spectra shape on the dimension of the system (the absence of hyperuniversality); the infinite value of the variance, calculated on the ground of the spectral density, which results in the fact that any natural system can blow up or get extinct in a finite time interval.

Neither of the SOC models considers fluctuations bounded. However, the features of any unbounded sequence strongly depend on the particular shape of its distribution. For example, the parameters S (total time-integrated amount of sliding) and $T$ (duration of an avalanche) in [10] vary independently of one another up to infinity that results in straightforward dependence of the power spectrum shape on their joint probability distribution.

On the contrary, a very important formal advantage of the boundedness is that the properties of the $1/f^\alpha$-type noise are independent of the particular distribution of the fluctuations [2]. The boundedness also ensures finite variance of the fluctuations that results in stable long-term evolution of the system.

In the fourth section it is found out that the considered fluctuations are embodied in the quantum spectra of the adsorption systems through the presence of a continuous band. It is superimposed on a discrete one that comes from the "rigid" part of the spectrum. The distinctive property of the continuous band is that it does *not* correspond to any real radiation. The shape of the continuous band is of the $1/f^\alpha$-type regardless to the features of the incident radiation and to the particularity of the system. Because of these properties the band is called alias $1/f^\alpha$-type noise.

Thus, the boundedness of fluctuations and the corresponding synchronisation are manifested both on the quantum and the macroscopic level through $1/f^\alpha$-type noise properties.

## 1. Randomly Disturbed Bounded Hamiltonian

The present paper is focused on the properties of the introduced above feedback, since it is the only implement that ensures long-term stable evolution eliminating the induced non-homogeneity. The latter is associated with the local diffusion-induced non-perturbative interactions available at all the surface reactions regardless to the particularity of the interface and the reaction mechanism. So, the problem whether the feedback is sensitive to the individuality of a system is of primary importance. The bearers of the individuality are the Hamiltonians that describe the adsorption and reaction potentials. Since the prerequisite of the feedback is the separation of the Hamiltonian into rigid and flexible parts, the task now is to establish that the separation always occurs under the non-correlated mobility of the adspecies and is it insensitive to the particularity of the unperturbed Hamiltonian. Furthermore, it will be proven that the flexible part becomes chaotic under non-correlated mobile disturbances.

The separation is associated with the stability to disturbances of a given Hamiltonian. Whenever a disturbance does not change significantly a state of its spectrum it is considered as a perturbation. This is justified for low-lying states only where the Hamiltonian is supposed to be stable to disturbances. In other words, the potentials that create the Hamiltonian are much stronger than the disturbances. On the contrary, even weak disturbances can cause essential changes of any highly excited state, since the highly excited states are weakly bounded. Thus, at the upper part of the spectrum the Hamiltonian becomes unstable to the disturbances. The instability implies that the disturbances should be considered in the same way as the creative potentials. Thus, any Hamiltonian separates into two parts: the "rigid" one that is associated with the unperturbed Hamiltonian and the "flexible" one that is associated with the disturbances.

The major source of disturbances to an adsorption Hamiltonian are the mobile adspecies and/or other local defects. They cause not only separation of a Hamiltonian but their non-correlated mobility makes the flexible parts that come from different sites non-identical even when the "rigid" parts are identical. Yet, the non-correlated mobility of the adspecies gives rise to an important for the feedback property shared by all the non-identical flexible parts. Further it is proven that the classical counterpart of any flexible part associated with permanent non-correlated bounded disturbances is chaotic. Bohigas et al conjecture states [5-6] that the nearest level spacing distribution of any Hamiltonian whose classical counterpart is chaotic is the Wigner one, namely:

$$P(E) = \frac{E}{S} \exp\left(-\frac{E^2}{S^2}\right) \qquad (1)$$

where $E$ is the spacing between the nearest levels, $S$ depends on the phase space volume of the Hamiltonian. Consequently, any state of any chaotic part is determined by a single parameter, namely by its energy. The insensitivity to the chemical identity of the species gives rise to the property that the any collision at the chaotic states neither requires specific local configuration nor introduces spatial correlations among colliding species. This property allows to consider the subsystem of species in the chaotic states always spatially homogeneous. The property will be utilized in the next section. Another important for the feedback property provided by the chaoticity will be presented later.

Our next step is to prove that the classical counterpart of any "flexible" part that comes from the influence of the non-correlated mobile species is chaotic. This is carried out by the use of the classical equations of motion:

$$\frac{dq_i}{dt} = \frac{\partial H}{\partial p_i} \qquad (2)$$

$$\frac{dp_i}{dt} = -\frac{\partial H}{\partial q_i}$$

where $H = \sum_{i}^{N(t)} V_i$ is the " flexible" Hamiltonian; $q_i, p_i$ $(i = 1,...,g)$ are the canonical variables and their conjugates; $g$ - the degrees of freedom and $V_i$ are the disturbances induced by the mobile species. Any "flexible" Hamiltonian $H = \sum_{i}^{N(t)} V_i$ has two crucial for the chaoticity properties: (I) each $\frac{\partial H}{\partial p_i}, \frac{\partial H}{\partial q_i}$ ($\forall i, j$) varies irregularly in the time course since each $H$ comprises terms that come from the non-correlated mobile adspecies and other local defects that are in the immediate neighborhood of the corresponding site; (ii) the variations of each $\frac{\partial V_i}{\partial q_j}, \frac{\partial V_i}{\partial p_j}$ ($\forall i, j$ at any instant) and of each $\frac{\partial H}{\partial p_i}, \frac{\partial H}{\partial q_i}$ are bounded. The range of variations is determined so that neither any term nor their sum destroys the Hamiltonian. The boundedness of the irregular sequence that constitutes the r.h.s. of eq.(2) is the major property of the type of equations subject to our previous papers [1-2]. It has been proven there that any its solution has chaotic properties. More precisely it has been established [2] that: a) any trajectory in the phase space is confined in a non-homogeneous strange attractor whose correlation dimension is non-integer; b) the Kolmogorov entropy is finite; c) the power spectrum is a continuous band of the shape $1/f^{\alpha(f)}$ where $\alpha(f) \to 1$ as $f \to 1/T$ ($T$ is the length of the sequence) and $\alpha(f)$ monotonically increases to the value $p > 2$ when the frequency approaches infinity. Thus, the solution of eq.(2) has chaotic properties regardless to the particularities of the disturbances sequence in the Hamiltonian $H = \sum_{i}^{N(t)} V_i$. This, in turn, justifies that the quantum "flexible" Hamiltonian is chaotic. Therefore, though the "flexible" parts of the Hamiltonians at different adsorption sites are not identical, their spectra have the same nearest level spacing distribution: the Wigner one (eq.(1))

Furthermore, the permanent boundedness of each $V_i$ and their sum ensures that the spectrum of the perturbed "rigid" part is always discrete. This immediately follows from the proof [11] that any self-adjoint initial Hamiltonian preserves that property under random bounded perturbation(s).

An immediate result of the above considerations is the insensitivity of the chaotic properties not only to the particularities of Hamiltonian but to its dimensionality as well because the chaotic features of the classical counterpart does not depend on $g$ in eq.(2). Moreover, the chaoticity does not depend on whether the Hamiltonian is attractive or repulsive.

The formation of bounded states over a repulsive Hamiltonian justifies our notion of a diffusion-induced non-perturbative interaction since it allows a temporary restrain of two species at the same adsorption site. It should be stressed that one of the species is always in the "rigid" part of the spectrum and thus it can relax to the ground state while the other one is in a chaotic state and it is allowed only either to migrate to another site or to desorb back to the gas phase.

Likewise, the notion of trapped species becomes rigorous: a trapped species is a species in a chaotic state.

The next task is to find out the influence of the collisions among the species in chaotic states on the feedback. Because of their fast non-correlated mobility, the rate of collisions between them is considerable. Next, it is put forth that the presence of any extra species at a single site changes the corresponding chaotic Hamiltonian. This immediately drives the feedback since it induces a non-radiative transition that dissipates via excitation of appropriate local cooperative modes. In turn, the latter immediately appears as a disturbance to the chaotic Hamiltonian producing non-perturbative changes in it. Thus, the elastic collisions between species in chaotic states is the weakest process that drives the feedback. In the next section it will be proven that this driving "force" is enough for the successful accomplishment of the synchronisation.

The above consideration makes the difference between our approach and the weak-coupling one apparent. Both approaches consider a collision between "chaotic" species perturbation. Yet, in the frame of the weak-coupling approach the effect of the cooperative excitations on the adsorption Hamiltonian is always considered a perturbation. The latter is not large enough to induce another transition and the process stops. The next transition happens only when other species collide brought together by their mobility. On the contrary, in our approach the cooperative modes contribute to the chaotic Hamiltonian changing it *non-perturbatively*. The latter results in inducing a new transition that dissipates again through the excitement of a local cooperative mode(s). In turn, it changes again the chaotic Hamiltonian non-pertirbatively inducing a new transition and so on.

Since the "chaotic" species are much less weakly bounded that the adspecies, it is plausible to consider their influence on a chaotic Hamiltonian as a perturbation. The rate of a transition caused by an elastic collision between "chaotic" species reads:

$$P = \frac{|V_{ji}|^2}{\hbar^2 \omega_{ij}^2} \qquad (3)$$

where $\langle j|V|i \rangle$ is the matrix element of the transition from the initial state $i$ to the final state $j$ under the perturbation $V$; $\hbar\omega_{ij} = E_j - E_i$, $E_i, E_j$ are the energies of the initial and final state correspondingly. The transition rate is determined under the consideration of a perturbation as a sudden inclusion [12], since the direct collision is active in a very short time.

The same nearest level spacing distribution (1) shared by the chaotic Hamiltonians that comes from different adsorption sites renders that the transition energies $\hbar\omega_{ij}$ are equiprobable in the range $[0, S]$. The equiprobability arises because the colliding "chaotic" species come from different adsorption sites. The range $[0, S]$ is set on the limit of $\hbar\omega_{ij}$ imposed by the phase space volume of the chaotic part which is of the order of $S$. On the other hand, the range $[0, S]$ determines the types of available cooperative excitations through which the feedback proceeds. Since $S$ is associated with the weakest bound states (the chaotic ones), it is to be expected that it is very small. Therefore the acoustic phonons certainly are an available type of cooperative excitations.

Since the local excitations comprise collective modes, any excitation is spread throughout the entire system. However, the effect of any of them is pronounced only up to its wave-length distance because the excitations at different sites have different frequencies and are not coherent. Evidently, the dissipation through cooperative excitations makes the feedback non-local: the feedback "response" covers the wave-length distance while a transition happens at a given site, i.e. at a point. Thus, the coupling among neighbor local fluctuations is carried out by the non-local feedback "response".

Very important for the further considerations is the particular case when the local excitations are acoustic phonons because their dispersion relation provides the following relation between the wave length and the frequency:

$$\lambda = \frac{c}{\omega_{ij}} \qquad (4)$$

where $c$ is the sound velocity. Eq. (4) provides the specific property of the coupling through acoustic phonons, namely the coupling covers larger areas when $\omega_{ij}$ decreases. Thus, smoothing out the energy difference between "chaotic" species results in spreading of the coupling over larger distances that eventually yields global synchronisation.

An important consequence of the separation of a bounded Hamiltonian into parts of excited and chaotic states is that the total number of states is always finite. Thus any relaxation or excitation can be completed in a *finite* time. On the contrary, when a Hamiltonian has an infinite number of states, the relaxation or excitation can take infinite time because the transition between any two states takes finite time no matter how small it can be.

## 2. Synchronisation

The synchronisation ensures the elimination of the induced spatial non-homogeneity via the spread of the coupling throughout the entire surface. Its hallmark is the evening of the individual adsorption rates. This immediately provokes the question about the extend to which the properties of the established global adsorption rate are associated with any property of an individual adsorption rate. The problem renders also the importance of finding out at what extend the spatio-temporal configuration of the local fluctuations is mapped onto the global adsorption rate. To the most surprise, it turns out that the global adsorption rate is always associated with certain individual adsorption rate. The successful selection of an individual adsorption rate is provided when initially the corresponding trapped species is in a certain local configuration whose features are independent of the details of the spatio-temporal configurations of the trapped species, excited ones and adspecies. Therefore, the gas phase participates to the global adsorption rate only through the average hitting rates of the reactants, i.e. through their partial pressures. Our task now is to find out explicitly how that feedback ensures the selection of a single adsorption rate so that the final adsorption rate is determined by its individual properties.

The implement of the synchronisation is the feedback that acts towards making the individual chaotic parts of the Hamiltonians identical. The non-local "response" of the chaotic Hamiltonians to a local perturbation provides coupling among neighbor fluctuations. The non-locality renders the coupled fluctuations "response" to the perturbation coherent. Thus, the spread of the coupling over larger and larger distance makes more and more chaotic Hamiltonians to act identically. The driver of the perturbations are the colliding trapped species that are at different states. According to eqs.(3) and (4) the feedback based on the excitation of acoustic phonons ensures an expansion of the coupling size ($\lambda$) among fluctuations when the difference in the states ($\omega_{ji}$) of the colliding trapped species decreases. Thus the gradual evening of the states results in spreading of the coupling over larger and larger area. An obvious outcome is that the local configuration of the smallest $\omega_{ji}$ "couples" the most neighbor chaotic Hamiltonians. Further, the largest coupling area provides the smallest changes of the state of every trapped species involved in that area. In turn, the smallest changes of $\omega_{ji}$ drives the most spread coupling. So, more and more trapped species "act" coherently with the species that initially is in the most favorable configuration. The process is completed when all the trapped species are in the same states of identical chaotic

Hamiltonians. This way, the global coupling makes all the species coherent and all the chaotic Hamiltonians identical. The successful selection of an individual adsorption rate is determined by the most favorable local configuration, i.e. the local configuration of the smallest $\omega_{ji}$. Its formation is a tricky interplay between two independent of one another processes: gas species non-coherent trapping and the adspecies non-correlated mobility. The lack of coherence among the trapping instants renders that initially the species are not at the same state but are distributed over the chaotic and the excited states. By undergoing a diffusion-induced non-perturbative interaction the local configuration around any species can be changed so that it may become the most favorable one. So, the independence of one another of the gas species trapping and adspecies mobility ensures insensitivity of the most favorable configuration to the particularities of the spatio-temporal configuration of the trapped species, excited ones and adspecies. At the same time the undergoing of a diffusion-induced non-perturbative interaction provides the most pronounced differences among the individual adsorption rates because it causes a transition from an excited to a chaotic state.

The problem that follows is whether the synchronisation itself imposes some specific properties to the global adsorption rate so that to blur the particularity of the "most favorable" individual adsorption rate. As far as the feedback is based on the excitation of acoustic phonons, the coupling is driven by a single parameter $\omega_{ij}$ (see eqs.(3)-(4)) regardless to the current scale. The natural limits of that feedback introduce the following scale: the boundedness of $\omega_{ij}$ in the range $\left[0, \frac{S}{\hbar}\right]$ determines the smallest coupling size, namely $\lambda_{cr} = c\hbar/S$. It is obvious that whenever the following relation between the size of the surface $L$, the adsorption site spacing $a$ and $\lambda_{cr}$:

$$a \ll \lambda_{cr} \ll L \tag{5}$$

holds, the coarse-grained spatial evolution of the feedback is scale-free. Further it is proven that its coarse-grained temporal evolution is also scale-free. The lack of specific to the synchronisation scale(s) ensures that the coarse-grained final adsorption rate is determined merely by the individual adsorption rate that comes from the "most favorable" local configuration.

The coherent non-local "response" of the feedback to any local perturbation implies the approximation of the area of that "response" as a "temporary" cluster. The premise of the notion "temporary cluster" is that all the species of the cluster are synchronised, i.e. they respond coherently to the current perturbation. However, the states at which different temporary clusters are synchronised are different. Then, the global synchronisation evolves through coupling between adjacent temporary clusters. The result of the coupling is the synchronisation of the interacting temporary clusters so that the one in the most favorable configuration imposes its individual adsorption rate onto the others. Since the statistical weight of the individual adsorption rate of any cluster is proportional to its area, the global synchronisation is completed when one cluster covers the entire available area. The coupling between any two adjacent clusters is constituted by the collisions among their species. The intensity of each collision is determined by (3) where $\hbar\omega_{ji} = E_j - E_i$, $E_i$ ($E_j$) is the energy of any species of the cluster $i$ ($j$). Due to the finite mobility rate of the trapped species the collision rate between species of different energies is determined by a strip around their boundary. The trapped species collisions are collisions at the chaotic states and thus they neither require nor introduce any specific local configurations. Consequently, the collision rate of interaction between adjacent temporary clusters does not involve any specific scale associated with the details of the trapped species interactions. Formally this is carried out by associating the collision rate with the power $X^{\nu(X)}$, where $X$ is a size of a given temporary cluster. $\nu(X)$ is a non-constant exponent determined so that $X^{\nu(X)}$ gives the collision strip "weight" expressed trough the current size of a the given temporary cluster. The advantage of that presentation is that the collision rate is scale-free. Further it is assumed that the mobility rate of the trapped species is constant during the synchronisation. This immediately renders the width of the collision strip constant. Hence, it is obvious that the gradual increase of the cluster size is accompanied by a gradual decrease of $\nu(X)$. The justification of the $\nu(X)$ limits is discussed further.

The premise that initially the species are randomly distributed over the chaotic and excited states renders all $\omega_{ij}$ equiprobable. Farther the equiprobability of $\omega_{ij}$ is sustained by the sharing of the same nearest level spacing distribution of all chaotic Hamiltonians. Then the initial iterations of the synchronisation of $N$ ($N$ arbitrary) temporary clusters are governed by the following system of ordinary differential equations:

$$\frac{dX_i}{dt} = \sum_{j=1}^{Z_i} \frac{|V_{ji}|^2}{\hbar^2 \omega_{ji}^2} X_i^{\nu(X_i)} \operatorname{sgn}(i,j) \tag{6a}$$

$$\frac{dX_j}{dt} = \sum_{k=1}^{Z_j} \frac{|V_{jk}|^2}{\hbar^2 \omega_{jk}^2} X_j^{\nu(X_j)} \operatorname{sgn}(j,k) \tag{6b}$$

..............................................

$$\frac{dX_N}{dt} = \sum_{l=1}^{Z_N} \frac{|V_{lN}|^2}{\hbar^2 \omega_{lN}^2} X_N^{\nu(X_{Nj})} \operatorname{sgn}(l,N) \tag{6c}$$

The initial conditions are:
$$X_i = X_j = ... = X_N = M \tag{7}$$

There is an additional condition, namely:
$$\sum_{i=1}^{N} X_i(t) = \sum_{i=1}^{N} X_i(0) = NM \tag{8}$$

In order to eliminate the dimension factor we rescale all $X_i$ by the factor $NM$. So, $X_i$ becomes $\frac{X_i}{NM}$. Thus, the range of rescaled variables is always $(0,1]$. The time is also rescaled with respect to the overall time of the synchronisation and also varies in the range $[0,1]$.

Next all $V_{ij}$ $(i,j = 1,...,N)$ are supposed equal because they come from the same type of interaction and so they are of the same order. Since $\omega_{ij}$ is the same for the $i-th$ and $j-th$ cluster, additional considerations are necessary to determine the sign of each term that involve these clusters. Let us consider each of the equations in (6) separately from one another, i.e. to consider all the terms in the r.h.s. positive. The cluster that grows fastest is that of the smallest $\sum_{j=1}^{z_i} \omega_{ji}$. For the sake of simplicity let it be the $i-th$ cluster. Figuratively, it means that the $i-th$ cluster "swallows up" all its neighbors. Then the contribution of the interaction between the $i-th$ and the $j-th$ cluster is positive for the $i-th$ cluster and is negative for the $j-th$ cluster. Thus, in this particular case, $\operatorname{sgn}(i,j)$ in (6a) is positive while $\operatorname{sgn}(j,i)$ in (6b) is negative.

Eqs.(6) has a single stable solution that is:

$$X_i = 1, \qquad (9a)$$
$$X_j = 0 \quad j \neq i \quad j = 1,\ldots,N \qquad (9b)$$

where $X_i$ is that cluster which is initially at the most favorable configuration. Eqs.(6) are unstable to arbitrarily small perturbations provided several clusters have the same rates at a given instant. The cluster that becomes in more favorable configuration under the perturbations "swallows up" the others.

An important property of eqs.(6) is that its solution does not depend on $N$. This fact allows to consider the synchronisation of an infinite system in the following way. Initially a partition of the system into areas of finite size is made. The above considerations are applied to each area. The next iteration involves interactions between new (larger) areas at which the previous areas appear as temporary clusters.

The initial condition implies that the trapped species of different energies are homogeneously distributed throughout the surface, i.e. they form a perfect mixture. It is plausible to suppose that initially the width of the collision strip and the cluster size are of the same order. This immediately sets the small size limit of $\nu(X)$, namely $\nu(X) = 1$. It was shown above that the enlargement of the clusters is accompanied by a gradual decrease of $\nu(X)$. The monotonic decrease of $\nu(X)$ in eqs.(6) provides permanently accelerated growing of the "most favorable" cluster. Its asymptotic reads:

$$X(t) \propto t^{\frac{1}{1-\nu(t)}} \qquad (10)$$

The integration of a power function with non-constant exponent is presented in the Appendix. The properties of $\nu(t)$ are determined by the diffeomorfism between $\nu(t)$ and $\nu(X)$. That diffeomorfism provides the scale-free behavior of the solution of eqs.(6). So, the monotonic decrease of $\nu(t)$ ensures an accelerated growing rate of the "most favorable" cluster (see (10)). The increasing growing rate of the "most favorable" cluster is a premise that at certain instant it becomes infinite in the sense that its form becomes so irregular that it borders to every other cluster. Thus, the formation of the infinite cluster sets the large-size limit of $\nu(X)$, namely $\nu = 0$.

The increasing growing rate of the "most favourable" cluster ensures its stability to the temporary clusters induced in later moments. Indeed, according to eq.(10), even a more advantageous than the "most favorable" initial local configuration cannot overcome the difference in the size due to the time lag.

The lack of any scale in eq.(10) is a warrant that the synchronisation does not impose any specific scale(s) onto the global adsorption rate. So, the latter is determined only by the individual properties of the "most favorable" one.

An important distinctive property of the above presented synchronisation is its insensitivity to the dimensionality of the system. Though a $2d$ case is considered, the dimensionality is neither explicitly nor implicitly involved in eqs.(6).

### 3. Size-Independence of the Macroscopic Fluctuations

The synchronisation is a temporary phenomenon since it lasts until the trapped species enter the rigid part of the spectrum. This, in turn constitutes a strong pulse-like behavior of the system temporal evolution due to the alternation of the synchronisation sessions and sessions of proceeding trough the rigid part of the spectrum. The property of the synchronisation to impose the "most favorable" individual adsorption rate onto the whole system gives rise to fluctuations of the global adsorption rate in the course of the time. Indeed, in the previous section it was shown that the most pronounced differences among the individual adsorption rates come from undergoing of a diffusion-induced non-perturbative interaction. However, in the Introduction it was presented that the adsorption rate of a species that undergoes diffusion-induced non-perturbative interaction is a multi-valued function. Thus, the particularity of the "most favorable" individual adsorption rate is determined by the established selection. The independence of one another between the trapping moments and adspecies mobility renders all the selections equiprobable. So, each of them can be involved equiprobably in the "most favorable" local configuration. Thus, at the next synchronisation session another selection of the adsorption rate happens to be the "most favorable" one. Then the adsorption rate temporal variations are constituted by the permanent random choice of one selection among all available. The range of variations is determined by the size of the largest fluctuations, i.e. the size of the selections. It is worth noting that the selections are specific to the system and each of them can be computed by an appropriate quantum-mechanical approach. Yet, the probability for undergoing of a single diffusion-induced non-perturbative interaction is proportional to the adspecies coverage and is insensitive to the details of the selections. This immediately makes the size of the individual adsorption rates proportional to the adspecies coverage. Since the global adsorption rate is always associated with some individual one, the macroscopic fluctuation size is the product of the size of an individual adsorption rate and the surface area. Thus, the macroscopic fluctuation size is proportional to the adspecies number.

It should be stressed that the amount of adspecies that the surface can adsorb is limited not by its size but by its saturation threshold. The latter, however, is set on the thresholds of the stability of the system that fix the limits of the energy and/or matter fluctuations so that the system remains permanently stable. This immediately renders the independence of the fluctuations limits from the size of the surface. An obvious property of the considered fluctuations is their boundedness: locally it is ensured by the property that only one species can be adsorbed at a single site; globally it is ensured by the presence of a saturation threshold. Previously, we have introduced [1-2] the idea that a stable evolution of any natural system is sustained if and only if the energy and matter fluctuations are bounded so that the system permanently remains within the thresholds of stability. Thus, the hallmark of a stable evolution of any natural system is the boundedness of any time series that depicts its temporal behaviour. It has been proven [2] that each irregular bounded sequence (BIS) has certain generic properties called $1/f^\alpha$-type noise. Hence, the considered in the present paper inevitable fluctuations should be regarded as the origin of the $1/f^\alpha$-type noise behaviour at the surface reactions. Indeed, the $1/f^\alpha$-type noise behaviour has been established at the catalytic oxidation of $HCOOH$ over supported $Pd$ catalyst [7].

The size-independence of the present fluctuations does not oppose them to the self-organisation of the spatio-temporal patterns. On the contrary, they are complementary events in the sense that they have different origin and appear at different types of sessions. The size-independent fluctuations appear at synchronisation sessions and their existence is the result of a tricky interplay between two independent from one another types of shot noise, namely non-coherent trapping of the gas species and non-correlated mobility of the adspecies. On the other hand, the self-organisation of spatio-temporal patterns is available at the "rigid" sessions where no such interplay takes place because all the trapped species enter coherently each "rigid" session. It is worth noting that the weak-coupling assumption is relevant at the "rigid" sessions.

# 4. Alias $1/f^{\alpha}$-type Noise

So far, the revelation of the boundedness and the synchronisation on the macroscopic level is considered. Yet, being a quantum phenomenon, it is to be expected that the synchronisation is manifested on the quantum level as well. Next, the issue how the synchronisation is related to the quantum spectra such as IR, EXAFS etc. is considered. At first sight the issue is controversial since one of the major assumptions introduced in the present paper is that the transitions in the chaotic part of the spectrum are non-radiative and their energy dissipates through excitation of appropriate cooperative modes such as acoustic phonons. Yet, the problem about the influence of the radiation on the synchronisation stands. Our first step is to verify that the synchronisation is "transparent" to the radiation regardless to its frequency. Then, the transitions in the chaotic part of the spectrum are not "mapped" in the corresponding quantum spectra. Thus, another feature of the synchronisation helps its revelation in the quantum spectra.

Locally any photon can interact with a species in a chaotic state if available. However, the outcome of these interactions is insignificant because: (I) the changes of the photon energy are negligible since the energies of the chaotic states are much smaller than the photon energies; (ii) these interactions does not affect synchronisation since the radiation is a gas of photons and thus cannot sustain any cooperative behavior. So, the radiation merely "rearranges" non-coherently the current energies of the trapped species. However, this does not modify fundamentally the synchronisation since the only successful way for its driving (the dissipation through excitation of cooperative modes) remains unaffected. Outlining, the radiation does not "interact" significantly with the synchronisation and elapses unchanged through the chaotic part of the spectrum. Therefore, the emitted radiation comes from the interaction between the incident photons and excited species at the rigid part of the spectrum. However, the temporal behaviour of the photon emission is fundamentally influenced by the strong alternation in the time course of synchronisation sessions and sessions of proceeding trough the rigid part of the spectrum. Indeed, since the synchronisation does not interact with any radiation, the photon emission is available only when the species are the rigid part of the spectrum. Thus the photon emission is not a continuous in the time process but exhibits a pulse-like behaviour: the subsequent emission sessions are separated by "refractory" intervals, i.e. the time intervals in which the emitted radiation elapses through a synchronisation session. Inasmuch as the length of the refractory intervals is much smaller than the length of the emission sessions it is plausible to assume that the successively emitted wave trains interfere. Evidently, that interference is embodied in the quantum spectra. Next, it is proven that the interference gives rise to a continuous band of certain shape superimposed on the discrete one that comes from the rigid part of the spectrum. It should be stressed that the continuous band does *not* correspond to any real radiation but it is a result of the pulse-like behaviour of the photon emission. Since this behaviour is an outcome of the properties of the synchronisation, the appearance of the continuous band serves as an indirect evidence for its existence.

We start with the consideration how the interference is embodied in the autocorrelation function of the photon emission. Hereafter $\tau$ denotes the length of an emitted wave train and $r_i$ is the length of the $i-th$ refractory interval. The length of the refractory intervals exhibits temporal variations because the synchronisation sessions are not identical. At steady external constraints the refractory intervals are uniformly distributed over a bounded range rendered by the shortest and largest duration of the synchronisation sessions ($r_{min}$ and $r_{max}$ correspondingly). For the sake of simplicity let us start with the case when only nearest wave trains interfere. Then the autocorrelation function $G(T)$ of the interference reads:

$$G(T) = \sum_{j=1}^{K} \sum_{n=1}^{N} |g(\omega_n)|^2 \exp(i\omega_n r_j) \tag{11}$$

where $T$ is the duration of the measurement, $K$ is the number of the refractory periods in the time series; $g(\omega_n)$ is the wave function amplitude of an emitted photon of energy $\hbar\omega_n$. Since the emitted photons comes from the rigid part of the spectrum where the nearest level spacing is always non-zero, the cross-section of any interaction (scattering) between a photon and an excited species is always finite regardless to the particularity of the interaction (scattering). Boundedness of the cross section renders $g(\omega_n)$ finite. In turn this renders $G(T)$ a bounded irregular function. Irregularity comes from the variation of the refractory interval lengths. Thus, being a BIS, the autocorrelation function shares the universal properties of the $1/f^\alpha$-type noise established in [2] and mentioned several times above. One of them is that the power spectrum of a time series of arbitrary but finite duration $T$ uniformly fits the shape:

$$S(f) \propto \frac{1}{f^{\alpha(f)}} \tag{12}$$

where $\alpha(f)$ monotonically increases from $\alpha\left(\frac{1}{T}\right) = 1$ up to $\alpha(f) > 2$ as $f$ approaches infinity. A distinctive property of the spectrum is that it has an artificial infrared edge $f_{\min}$ associated with the length of a measurement (correspondingly time series) $T$ through the relation $f_{\min} = \frac{1}{T}$. It has been proven [2] that this edge is related to the boundedness of the time series itself and is insensitive to the particularity of the fluctuation succession.

A distinctive property of the pulse-like behaviour is that the presence of refractory intervals imposes limits on the number of subsequent wave trains that can interfere at an instant and thus provides the boundedness of the autocorrelation function $G(T)$. That number varies in the range $\left[1, \frac{\tau}{r_{\min}}\right]$. On the contrary, any random emission violates boundedness since the lack of refractory intervals allows the number of interfering wave trains to become arbitrarily large at any instant. Thus, the boundedness of the autocorrelation function is justified only for the pulse-like behaviour.

Summarising, the pulse-like behaviour of the photon emission is manifested through persistent appearance in the quantum spectra of continuous band of universal shape superimposed on the discrete one regardless to the features of the incident radiation. The appearance and the shape of the band are insensitive to the particular properties of both the incident radiation and the details of the studied system. Once again it is worth noting that the continuous band does *not* correspond to any real radiation. That is why the band is called alias $1/f^\alpha$-type noise.

It should be stressed that the $1/f^\alpha$-type behaviour of the quantum spectra predicted by us is fundamentally different from that mechanism of quantum $1/f$ noise proposed by Handel [13-14]. He has supposed that the $1/f$ noise comes from interference between elastically and inelastically scattered waves, which emerge when a beam of particles is scattered under the influence of a potential. The model predicts $1/f$ noise in any system whenever the cross section for scattering of particles exhibits an infrared divergence of low-frequency excitations. Later this mechanism has been strongly criticised [15 and references therein] and it has been rigorously proven [16] that it does not bring about the $1/f$ behaviour. Our considerations about the "transparency" of the synchronisation to the radiation are in accordance with this critic. Indeed, in the present model the low-frequency transitions that exhibit infrared divergences are non-radiative and thus are not "mapped" in the quantum spectra.

Next we present a simple discrimination criterion between Handel and our model. The major property of the $1/f^\alpha$-type power spectrum is the persistent presence of an infrared edge $\omega_{\min}$ that is inverse proportional to the length of a measurement $T$, namely $f_{\min} = \frac{1}{T}$. On the contrary, Handel has predicted that the theoretical power spectrum is $1/f$ for all the frequencies in the range $[0,\infty)$. However, this yields that the power spectrum coming from a measurement of a duration $T$ does not comprise any perceptible frequency associated with $T$. Indeed, let the predicted power spectrum is $g(f)$, $f \in [0,\infty)$. Then, the power spectrum that comes from a measurement of duration $T$ is:

$$g(f,T) = \frac{1}{T}\int_0^T S(t)\exp(ift)dt \tag{13}$$

where $S(t) = \int_{-\infty}^{\infty} g(f')\exp(-if't)df'$ \hfill (14)

Then:

$$g(f,T) = \frac{1}{T}\int_0^T \exp(ift)\int_{-\infty}^{\infty} g(f')\exp(-if't)df'dt =$$
$$= \int_{-\infty}^{\infty} \frac{\sin((f-f')T)}{(f-f')T} g(f')df' \tag{15}$$

It is obvious that $g(f,T)$ does not comprise any perceptible frequency associated with $T$. Furthermore, the following limit holds:
$\lim_{T\to\infty} g(f,T) = g(f)$ \hfill (16)

Thus, the increase of $T$ makes the experimental and theoretical power spectra closer and closer. In particular, (13)-(16) yield that the experimental power spectrum should have no infrared edge associated with $T$. Thus, the infrared edge dependence (or its lack) on the length of the measurement serves as a criterion for discrimination between predicted by us $1/f^\alpha$-type behaviour and the quantum $1/f$ noise predicted by Handel.

**Conclusions**

The general necessary condition providing a long-term stability of any system requires persistent presence of a coupling mechanism that makes the local fluctuations behavior coherent. Otherwise, the system would blow up or get extinct under the permanent development of non-correlated and unbounded local fluctuations.

The boundedness of fluctuations makes the developed so far approaches such as Fokker-Planck equation, master equation and approaches used Wiener process for modelling noise sources inappropriate to the problem. Moreover, the inevitable presence of the diffusion-induced noise at surface reactions renders the widespread weak-coupling approach to the adspecies-interface interaction out of place as well.

That is why we introduce two major assumptions that are the grounds of the successful coupling mechanism. The first one is that any bounded adsorption Hamiltonian separates into a "rigid" and a "chaotic" part under the disturbances induced by the non-correlated mobility of the adspecies. The distinctive property of the "chaotic" parts is their sensitivity to the environment. Therefore, the dissipation of any transition energy excites certain cooperative modes that in turn participate to the "chaotic" part of the corresponding Hamiltonian. This causes a change of the "chaotic" part of the Hamiltonian that induces a new transition. Our next assumption is that the induced transitions are non-radiative and always dissipate through excitation of appropriate cooperative modes. Thus, our two assumptions give rise to a non-perturbative feedback between the trapped species and the surface whose most prominent property is that it "works" permanently until the global "response" to further perturbations becomes coherent.

It is proven that the weakest coupling mechanism, that driven by the collisions between species in chaotic states, has universal properties. It is insensitive neither to the chemical identity of the reaction species, nor the properties of the reaction mechanism, surface itself and the details of adspecies configuration.

It is a scale-free process that renders the established finally global adsorption rate always associated with that individual one which initially is in the most favourable local configuration. That is why the process of coupling among local fluctuations is called synchronisation. The independence from one another of the trapping moments and from the adspecies mobility renders that after undergoing a diffusion-induced non-perturbative interaction any trapped species may occur in the most favourable local configuration regardless to the details of the adspecies spatio-temporal configuration. The scale-free behaviour of eqs.(6) provides also that the synchronisation itself does not introduce any specific scale to the "most favourable" adsorption rate. Thus, the global adsorption rate follows the major features of an individual one. One of them is that the individual adsorption rate is a multi-valued function. The independence from one another of the trapping moments and from the adspecies mobility makes all its selections equiprobable. Then, the global adsorption rate is also a multi-valued function and the established selection is randomly chosen among all available ones.

The separation of the adsorption Hamiltonians into "chaotic" and "rigid" parts strongly affects the temporal behaviour of the system. The alternation of sessions of synchronisation and sessions of proceeding through rigid parts makes the temporal behaviour pulse-like. As a result, the global adsorption rate varies with the synchronisation sessions since it is established through random choice of one selection among all available. The distinctive property of the global adsorption rate variations is their boundedness imposed by the saturation threshold of the system. The boundedness itself is the hallmark of a stable long-term evolution of any natural system regardless to the details of the fluctuation series. So, on macroscopic level it is manifested through the $1/f^{\alpha}$ -type noise properties of the time series that depict the temporal behavior of the system.

Another important property of the established in the present paper fluctuations is the independence of their amplitude from the surface size. However, this property does not oppose them to the self-organisation of the spatio-temporal patterns. On the contrary, they are complementary events in the sense that they have different origin and appear at different types of sessions. The size-independent fluctuations appear at synchronisation sessions and their existence is the result of a tricky interplay between two independent from one another types of shot noise: non-coherent trapping of the gas species and the non-correlated mobility of the adspecies. On the other hand, the self-organisation of spatio-temporal patterns is available at the " rigid" sessions where no such interplay takes place because all the trapped species enter coherently each "rigid" session. It is worth noting that the weak-coupling assumption is relevant at the "rigid" sessions.

The pulse-like temporal behaviour is manifested also in the quantum spectra such as IR, EXAFS etc. Since the synchronisation is "transparent" to the radiation, the permanent alternation of the "synchronisation" and "rigid" sessions give rise to interference among the wave trains emitted at subsequent "rigid" sessions. The important distinctive properties of the autocorrelation function of that it is a bounded irregular function regardless to the particularities of the surface reaction and the radiation. Then, it is embodied in the quantum spectra as a persistent continuous band of the shape $1/f^{\alpha(f)}$ superimposed to a discrete one that comes from the "rigid" part of the spectrum.

Apart from the shape, the other distinctive property that is typical for the $1/f^{\alpha}$ -type noise is the presence of infrared edge associated with the length of the measurement through the relation $f_{\min} = \dfrac{1}{T}$.

These properties together with the fact that the band does not correspond to any real radiation elucidate its name: alias $1/f^{\alpha}$ -type noise.

Thus, the boundedness of fluctuations and the corresponding synchronisation are manifested both on quantum and macroscopic level through $1/f^{\alpha}$ -type noise properties.

# Appendix

The task of the appendix is to find out how to integrate a power function with a non-constant exponent. So, the next item is the calculation of:

$$J(a,b) = \int_a^b x^{\pm v(x)} dx. \tag{A.1}$$

where $v(x)$ is a continuous function with finite everywhere derivatives. First we consider the case when $v(x) \neq 1$ for any $x \in [a,b]$. The interval $[a,b]$ is divided into subintervals of equal length $\varepsilon$ and the value of $v(x)$ is set constant equal to its value at the lower limit of any subinterval. Then:

$$J(a,b) = \frac{b^{1 \pm v(b)}}{1 \pm v(b)} - \frac{a^{1 \pm v(a)}}{1 \pm v(a)} + \lim_{\varepsilon \to 0} \sum_{k=1}^{[(b-a)/\varepsilon]} R_k, \tag{A.2}$$

where:

$$R_k = (a+k\varepsilon)^{\pm v(a+(k-1)\varepsilon)+1} - (a+k\varepsilon)^{\pm v(a+k\varepsilon)+1} \approx$$
$$\approx (a+k\varepsilon)^{\pm v(a+k\varepsilon)+1}\left((a+k\varepsilon)^{-\delta_k \varepsilon} - 1\right),$$

where:

$$\delta_k = v'(a+k\varepsilon).$$

Since it is supposed that $v(x)$ has finite everywhere derivatives, $\delta_k$ is finite for every $k$. In turn, this provides that $\left((a+k\varepsilon)^{-\delta_k \varepsilon} - 1\right) \to 0$ at $\varepsilon \to 0$. So each term $R_k = 0$ at $\varepsilon = 0$. Thus:

$$J(a,b) = \frac{b^{1 \pm v(b)}}{1 \pm v(b)} - \frac{a^{1 \pm v(a)}}{1 \pm v(a)}. \tag{A.3}$$

The justification that for a scale free process $v(x)$ is a continuous linear function with a finite derivative is presented in [17].

Let us now consider the case when $v(x)$ is a continuous function with finite everywhere derivatives such that it crosses 1 at the point $x = c$ and $c \in [a,b]$. Then:

$$J(a,b) = \lim_{\varepsilon \to 0}\left[\int_a^{c-\varepsilon} x^{\pm v(x)} dx + \int_{c+\varepsilon}^b x^{\pm v(x)} dx\right] =$$
$$= \frac{b^{1 \pm v(b)}}{1 \pm v(b)} - \frac{a^{1 \pm v(a)}}{1 \pm v(a)} + \lim_{\varepsilon \to 0}\left((c-\varepsilon)^{\pm v(c-\varepsilon)+1} - (c+\varepsilon)^{\pm v(c+\varepsilon)+1}\right). \tag{A.4}$$

Taking into account that:

$$|v(c-\varepsilon)| = 1 - v'(c)\varepsilon$$
$$|v(c+\varepsilon)| = 1 + v'(c)\varepsilon,$$

$$\lim_{\varepsilon \to 0}\left((c-\varepsilon)^{\pm v(c-\varepsilon)+1} - (c+\varepsilon)^{\pm v(c+\varepsilon)+1}\right) \approx \lim_{\varepsilon \to 0}\left(c^{-v'(c)\varepsilon} - c^{v'(c)\varepsilon}\right) \to c^0 - c^0 = 0.$$

Therefore again:

$$J(a,b) = \int_a^b x^{\pm v(x)} dx = \frac{b^{1 \pm v(b)}}{1 \pm v(b)} - \frac{a^{1 \pm v(a)}}{1 \pm v(a)}. \tag{A.5}$$

However, our result seems controversial because the derivative of $J(a,b)$ deviaties from the l.h.s. of (A.5). Next a proof that this argument is alias and (A.5) holds whenever $v(x)$ is a continuous function with finite everywhere derivatives is presented.

To simplify the further calculations let us present (A.5) in a slightly modified form, namely:

$$J(a,b) = \int_a^b (\pm v(x)) x^{\pm v(x)-1} dx = b^{\pm v(b)} - a^{\pm v(a)} \quad (A.6)$$

Indeed, the application of the formal differentiating rules to the function $x^{\pm v(x)}$ in its exponential presentation:

$$x^{\pm v(x)} = \exp(\pm v(x) \ln(x)) \quad (A.7)$$

yields:

$$\left(x^{\pm v(x)}\right)' = \pm v(x) x^{\pm v(x)-1} + v'(x) \ln(x) x^{\pm v(x)} \quad (A.8)$$

Obviously (A.8) deviates from the l.h.s. of (A.6) by the factor $v'(x) \ln(x) x^{\pm v(x)}$.

Let us now apply the rigorous definition of a derivative. In our case it reads:

$$\left(x^{\pm v(x)}\right)' = \lim_{\varepsilon \to 0} \frac{(x+\varepsilon)^{\pm v(x+\varepsilon)} - x^{\pm v(x)}}{\varepsilon} =$$

$$\lim_{\varepsilon \to 0} \frac{\exp(\pm v(x+\varepsilon)\ln(x+\varepsilon)) - \exp(\pm v(x)\ln(x))}{\varepsilon} \quad (A.9)$$

For each $x$ fixed and non-zero one can define a small parameter $\frac{\varepsilon}{x}$. Hence, the Taylor expansion of $v(x+\varepsilon)\ln(x+\varepsilon)$ to the first order reads:

$$v(x+\varepsilon)\ln(x+\varepsilon) = (v(x) + \varepsilon v'(x))\left(\ln\left(1 + \frac{\varepsilon}{x}\right) + \ln(x)\right) =$$

$$= (v(x) + \varepsilon v'(x))\left(\frac{\varepsilon}{x} + \ln x\right) = v(x)\ln(x) + v(x)\frac{\varepsilon}{x} + \varepsilon v'(x)\ln(x) \quad (A.10)$$

Then:

$$\exp(\pm v(x+\varepsilon)\ln(x+\varepsilon)) = x^{\pm v(x)} \exp\left(\pm v(x)\frac{\varepsilon}{x}\right) \exp(\varepsilon v'(x)\ln(x)) \quad (A.11)$$

The expansion of $\exp\left(\pm v(x)\frac{\varepsilon}{x}\right)$ to the first order of $\frac{\varepsilon}{x}$ yields $\left(1 \pm v(x)\frac{\varepsilon}{x}\right)$. However, the expansion of the term $\exp(\varepsilon v'(x)\ln(x))$ into series is not unambiguous even when $\varepsilon \ln(x) \to 0$. This is because $\varepsilon$ and $x$ do not participate the Taylor series in a scaling invariant form. This renders the value of the derivative scale-dependent. Luckily, a scale-invariant form of the derivative is available by the use of the limit order: $\varepsilon \to 0$ faster than $x \to 0$. It makes the term $\varepsilon v'(x)\ln(x)$ equal to zero and correspondingly the term $\exp(\varepsilon v'(x)\ln(x))$ equal to 1 at each value of $x$. It is worth noting that the above limit order provides both the existence of the small scaling-invariant parameter $\frac{\varepsilon}{x}$ in the first exponential term and renders the limit 1 for the second exponential term in (A.11).

Thus, (A.8) yields:

$$\left(x^{\pm v(x)}\right)' = \pm v(x) x^{\pm v(x)-1} \quad (A.12)$$

So, (A.12) matches our result (A.6).